\documentclass[prl,aps,twocolumn,groupaddress]{revtex4-1}
\usepackage{latexsym}
\usepackage{graphicx}
\usepackage{verbatim}
\usepackage{multirow}
\usepackage{amsmath}
\usepackage{mathrsfs}
\usepackage{float}
\usepackage{feynmf}

\newcommand{\be}{\begin{equation}}
\newcommand{\ee}{\end{equation}}
\newcommand{\ba}{\begin{eqnarray}}
\newcommand{\ea}{\end{eqnarray}}
\newcommand{\baa}{\begin{eqnarray*}}
\newcommand{\eaa}{\end{eqnarray*}}

\def\be{\begin{equation}}
\def\ee{\end{equation}}
\def\bea{\begin{eqnarray}}
\def\eea{\end{eqnarray}}

\def\C60{A$_x$C$_{60}$}

\def\HgCu3{HgCa$_2$Cu$_3$O$_{8+y}$}
\def\HgCu4{HgBa$_2$Ca$_3$Cu$_4$O$_{10+y}$}
\def\TlCu{Tl$_2$Ba$_2$CuO$_{6+\delta}$}
\def\TlCu3{Tl$_2$Ba$_2$Ca$_2$Cu$_3$O$_{10+y}$}
\def\TlCu4{Tl$_2$Ba$_2$Ca$_3$Cu$_4$O$_{12+y}$}

\def\BiCu3{Bi$_2$Sr$_2$Ca$_{2}$Cu$_3$O$_y$}

\def\8LSCO{La$_{1.88}$Sr$_{.12}$CuO$_4$}
\def\110LNSCO{La$_{1.5}$Nd$_{0.4}$Sr$_{0.1}$CuO$_{4}$}
\def\stage4LCO{La$_{2}$CuO$_{4+\delta}$}
\def\Y248{YBa$_2$Cu$_4$O$_8$}

\def\NbSe2{NbSe$_2$}
\def\TaSe2{TaSe$_2$}
\def\TiSe2{TiSe$_2$}

\begin{document}
\title{Absence of nematic ordering transition in a diamond lattice: Application to $\mathrm{FeSc_2S_4}$}
 \author{Chandan Setty}
\affiliation{Department of Physics and Institute for Condensed Matter Theory, University of Illinois 1110 W. Green Street, Urbana, IL 61801, USA}
\author{Zhidong Leong}
\affiliation{Department of Physics and Institute for Condensed Matter Theory, University of Illinois 1110 W. Green Street, Urbana, IL 61801, USA}
\author{Shuyi Zhang}
\affiliation{Department of Physics and Institute for Condensed Matter Theory, University of Illinois 1110 W. Green Street, Urbana, IL 61801, USA}
\author{Philip W. Phillips}
\affiliation{Department of Physics and Institute for Condensed Matter Theory, University of Illinois 1110 W. Green Street, Urbana, IL 61801, USA}
\begin{abstract}
Recent neutron scattering observations by Plumb \textit{et al}. \cite{Broholm2016} reveal that the ground state of $\mathrm{FeSc_2S_4}$ is magnetic with two distinct Fe environments, instead of a quantum spin liquid as had been previously thought. Starting with the relevant $O(N)$-symmetric vector model of $\mathrm{FeSc_2S_4}$, we study how the discrete ($Z_2$) and continuous rotational symmetries are successively broken, yielding nematic and ordered phases.
At high temperatures, we find that the nematic order parameter falls as $T^{-\gamma}$ ($\gamma>0$), and therefore, $\mathrm{FeSc_2S_4}$ lacks any distinct nematic ordering temperature. This feature indicates that the three-dimensional diamond lattice of $\mathrm{FeSc_2S_4}$ is highly susceptible to the breaking of Ising symmetries, and explains the two distinct Fe environments that is present even at high temperatures, as seen by M\"{o}ssbauer and far infrared optical spectroscopy.
\end{abstract}
\maketitle
\textit{Introduction:} 
Frustrated magnetic systems, resisting ordering to the lowest temperatures, arise from an intricate  interplay between lattice geometry and the sign of the magnetic interactions.  While no single measurement can characterize the failure of a magnetic system to order, thereby remaining in a quantum spin liquid state, a distinct measure of the frustration is a large value of the ratio $f = |\Theta_{\rm cw}|/T_c$, where $\Theta_{cw}$ (proportional to the strength of the exchange interaction) is the Curie-Weiss temperature, and $T_c$ is the transition temperature; the system is considered frustrated in the regime $T_c<T<\Theta_{cw}$. In the class of materials $AB_2X_4$, which are known as spinels, the exchange interactions are frustrated because the A-site atoms form a diamond lattice and are surrounded tetrahedrally by the X-site atoms. Consequently, numerous papers have proposed that the ground state of these materials is of the spin liquid type \cite{Loidl2006-PRB,Balents2007-NatPhy,Vishwanath2008,Loidl2009-PRB,Nagler2011}. In particular, because the frustration parameter in FeSc$_{2}$S$_4$ is enormous ($f\approx 1000$), this material has risen to the fore \cite{Tsurkan2004,Loidl2004,Loidl2005-Vibronic,Broholm2012} as a leading candidate for a spinel exhibiting quantum spin liquid behaviour. 

However, the recent neutron scattering measurements by Plumb \textit{et al}. \cite{Broholm2016} are surprising, because they found that powdered samples of $\mathrm{FeSc_2S_4}$  exhibit a magnetic ordering transition at $11.8$ K. With $|\Theta_{cw}|\approx 45$ K \cite{Tsurkan2004}, this observation drastically reduces the frustration parameter in this material from a thousand to about $f\sim 4$. Their observations also uncovered a small and `incipient' cubic to tetragonal structural transition  ($c/a=0.998$) that closely accompanies the formation of orbital order; both of these phases precede the magnetic transition and continue to prevail even at high temperatures. The structural transition distorts the sulfur atoms coordinating the Fe ions, and in the process leaves the two Fe sublattices surrounded by inequivalent atomic potentials. In this new lattice environment with a lower symmetry, the hole in the A sublattice occupies the $d_{z^2}$ orbitals, while that in the B sublattice occupies the $d_{x^2-y^2}$ orbitals.

 In fact, the presence to two distinct Fe environments was present even in the original M\"ossbauer data\cite{Gibart1976,Kim2008} as noticed recently by  Broholm and collaborators\cite{Broholm2016}.  Additionally, far infrared optical absorption measurements \cite{Loidl2015,Haeuseler2002} detected two distinct bands near 467 cm$^{-1}$  up to 300 K, indicating a high-temperature symmetry broken phase. Thus, in contrast with previous reports\cite{Tsurkan2004,Loidl2004,Loidl2005-Vibronic,Broholm2012},  the authors \cite{Broholm2016} concluded that there is a strong indication of a phase with broken $Z_2$ sublattice symmetry, followed by the conventional regime in which continuous spin rotational symmetry is broken. 

It is this experimental puzzle that we address in this paper. 
Prior theoretical works on $\mathrm{FeSc_2S_4}$ have focused sharply \cite{BalentsNature,Schnyder2009,Balents2009,Balents2015} on the competition between spin-orbit and Kugel-Khomskii \cite{Khomskii1982} type exchange interactions, and have obtained a phase diagram containing a spin-orbit singlet phase and a magnetically/orbitally ordered phase separated by a quantum critical point (QCP). Consistent with existing experimental data \cite{Armitage2015,Loidl2015-PRB}, these works also argued that $\mathrm{FeSc_2S_4}$ lies close to the QCP on the spin-orbit singlet side of the phase diagram. The experiment of Plumb \textit{et al}. \cite{Broholm2016}, in contrast, shows that $\mathrm{FeSc_2S_4}$ lies on the magnetic side of this yet unobserved QCP.

In this work, using the order by disorder mechanism, we aim to provide a theoretical description of these nematic and ordered phases observed in $\mathrm{FeSc_2S_4}$. We begin by modeling the spins with an $O(N)$ symmetric vector model, where the spins are represented by $N$-component real vectors in three-dimensional space. Using the Hubbard-Stratonovich transformation, we decouple the biquadratic terms and define a generalized nematic order parameter in the context of the diamond lattice. We then study the temperature dependence of the spin nematic order parameter, and investigate the development of long-range magnetic order. 
In the large $N$ limit, we find that, contrary to a few possible models proposed in Ref. \onlinecite{Broholm2016}, the nearest neighbor (NN) and next-nearest neighbor (NNN) exchange interactions ($J_1$ and $J_2$ respectively) need to be comparable in order to fit experimental data. Moreover, the spin nematic order persists even at high temperatures; in the limit $T/J_1\gg1$, the nematic order falls as a power law proportional to $T^{-\gamma}$, $\gamma>0$. This indicates that the three-dimensional diamond lattice is highly susceptible to $Z_2$ symmetry breaking and explains the presence of two distinct Fe environments even at high temperatures, as seen by M\"{o}ssbauer \cite{Gibart1976,Kim2008} and far infrared optical spectroscopy \cite{Loidl2015,Haeuseler2002}. This is unlike the 2D case \cite{Kivelson2008,Kivelson2012} where there is a distinct transition with a discontinuity in the first derivative with temperature. The effects of including orbitals into the theory are detailed in the Supplementary Material.

\begin{figure}[h!]
\includegraphics[width=1.9in,height=1.8in]{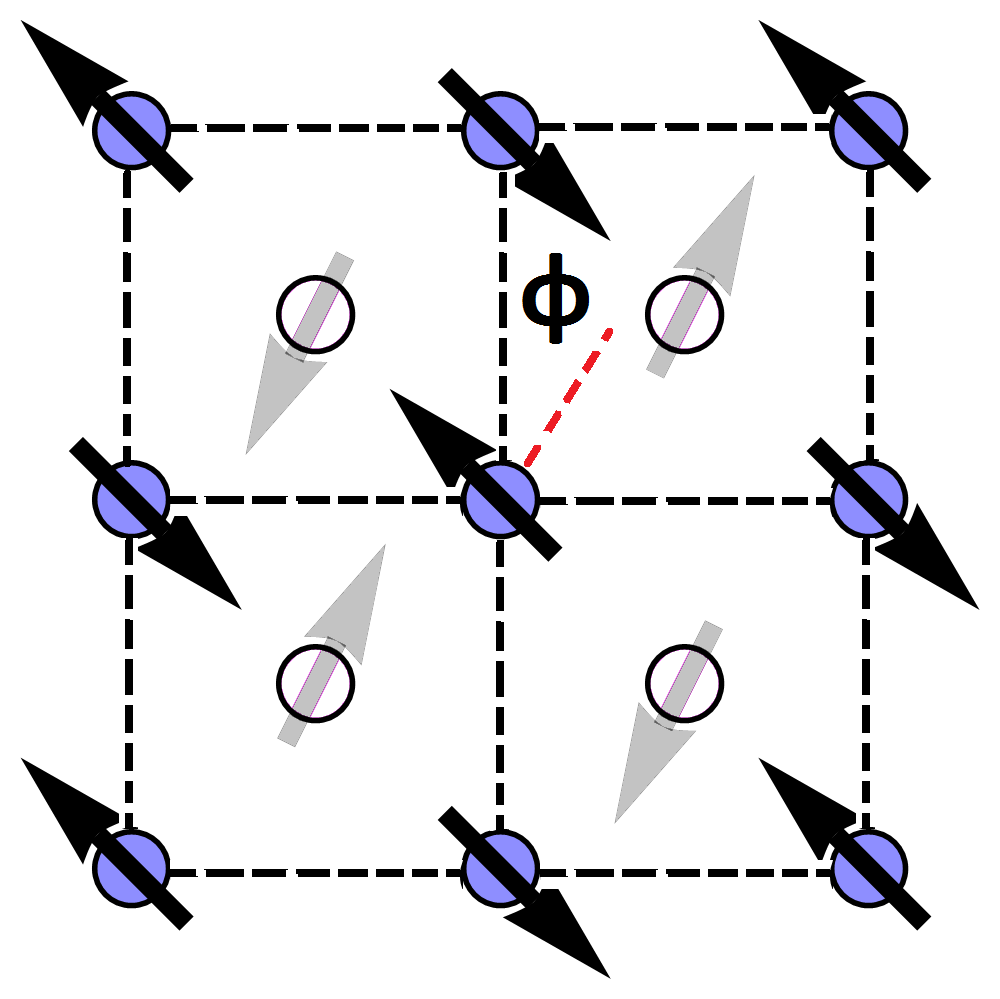}\hfill%
\caption{Two intercalated square lattices (solid and open circles) with antiferromagnetic order on each sublattice. Each atom in a sublattice either forms the center of a plaquette of the other sublattice, or could be displaced along the $c$ axis. The spins on one sublattice are oriented at an angle $\Phi$ with respect to the spins on the other sublattice.} \label{TwoSublattice}
\end{figure}

\textit{Order by disorder}: Apart from the conventional breaking of continuous spin rotational symmetries leading to ordered phases, Hamiltonians describing magnetic systems can also spontaneously break an additional discrete Ising ($Z_2$) symmetry associated with permutations of the sublattices \cite{Villain1980,Shender1982,Henley1989,Tsvelik2007}. This mechanism, widely referred to as the `order by disorder', has been extensively reported in high-temperature superconductors, such as the copper-based \cite{Sachdev1991,Brown2006} and iron-based superconductors \cite{Kivelson2008,Schmalian2014}. The key physics underlying this mechanism stems from biquadratic spin contributions \cite{Larkin1990} derived from integrating out short wavelength quantum fluctuations that are not initally present in the classical versions of the action. A representative system \cite{Tsvelik2007} where this is realized is the double layered antiferromagnet, schematically shown in Fig. \ref{TwoSublattice}. The emergent biquadratic terms break the continuous symmetry (and hence the degeneracy) with respect to arbitrary rotations (angle $\Phi$ in Fig. \ref{TwoSublattice}) between the sublattices. At the classical level, this symmetry exists even in the presence of inter-sublattice couplings. The net effect of the high energy quantum fluctuations on the classical action, then, is to lower the continuous rotational symmetry to a discrete Ising symmetry corresponding to a relative sublattice orientation of either 0 or $\pi$. Lowering the temperature can then break the order parameter symmetry space $O_j(N)\times Z_{2j}$ ($j=$ spin, orbital, etc) through successive phase transitions for each participating symmetry, thereby leading to nematic and/or ordered phases. Inspite of the rich potential that lies latent in these ideas, their applicability outside two-dimensional layered systems has been limited \cite{Paramekanti2010,Vishwanath2007}. It is, therefore, of great interest to further explore other classes of systems where similar physics can be realized in more general settings.

\begin{figure}[h!]
\includegraphics[width=1.7in,height=1.55in]{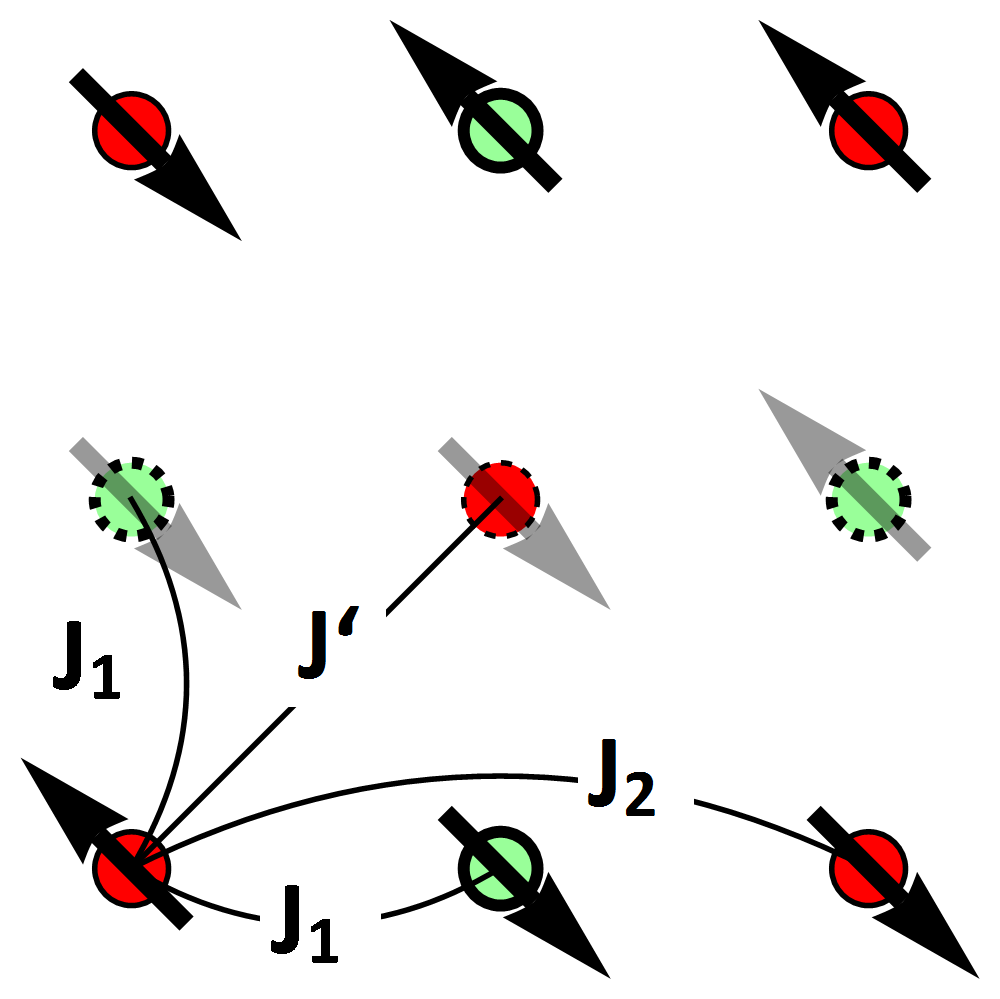}\hfill%
\includegraphics[width=1.7in,height=1.55in]{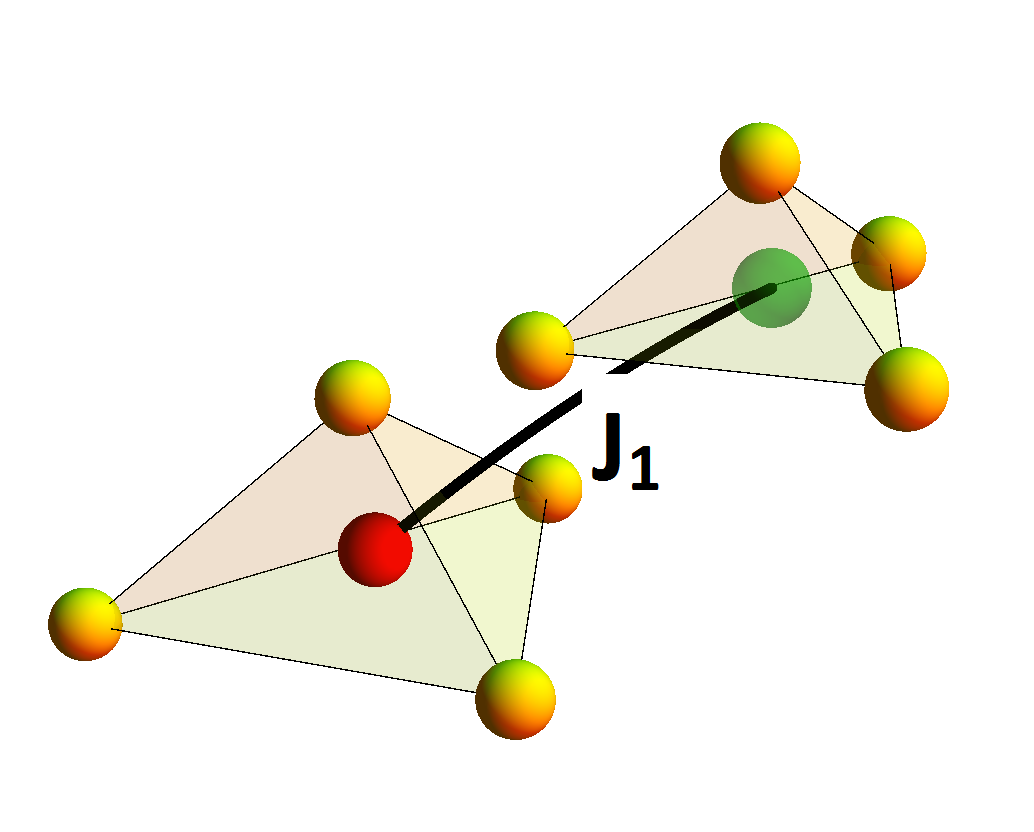}%
\caption{(Left) Magnetic structure proposed in \cite{Broholm2016} with exchange interactions defined. Red (dark) and green (light) disks denote the two Fe sublattices with the arrows pointing in the spin direction. The dark (light) arrows belong to the top (bottom) two layers. The disk boundaries order the various layers along the $c$-axis from the readers viewpoint --- (top to bottom) thick solid, thin solid, thick dashed and thin dashed. (Right) The sulfur tetrahedra surrounding each Fe sublattice. At lower temperatures, the tetrahedron about one of the Fe sublattice contracts and the other expands. }\label{DiamondMagnetic}
\end{figure}

\textit{Theory}: The partition function for the spin only degrees of freedom (the role of the orbital degree of freedom is presented in the Supplementary Material) is written as
\begin{eqnarray} \label{SpinPartitionFunction}
 \mathscr{Z}&=& \int \mathscr{D} \vec{\phi}_1 \mathscr{D} \vec{\phi}_2 \exp\left[ - \beta N \int d^3 \vec r \,\mathscr{L}\left(\phi_1^a(\vec r), \phi_2^a(\vec r) \right)\right]
\end{eqnarray}
where $\phi_1^a (\vec r), \phi_2^a (\vec r)$ are the $a$th components of the $O(N)$ vector on sublattices $j=1,2$ at lattice site $\vec r$. For simplicity, we will henceforth supress the index $a$ on $\phi_j(\vec r)$, keeping in mind that they refer to the individual components of a vector. We also denote $\mathscr{L}$ as the Lagrangian density, $N$ as the number of spin components, and $\beta$ as the inverse temperature. Defining $J_1$ and $J_2$ to be the NN and NNN magnetic exchange couplings, respectively, we can write the Lagrangian, $\mathscr{L}$, in the continuum limit as
\begin{eqnarray} \nonumber
\mathscr{L}\left(\phi_1, \phi_2 \right) &=&  \frac{J_2}{2} \sum_{\substack{j=1,2\\ i=x,y,z}}\left(\partial_{i} \phi_j (\vec r)\right)^2 - N K_{\phi} \left( \phi_1 (\vec r) \phi_2 (\vec r)\right)^2 \\
&&+ J_1 \sum_{\vec a_{\mu}} \partial_{\vec a_{\mu}} \phi_1(\vec r) \hspace{0.15cm} \partial_{\vec a_{\mu}} \phi_2(\vec r).\label{SpinLagrangian}
\end{eqnarray}
We note that the coupling constants $J_1$ and $J_2$ contain factors proportional to the magnitude of the spin angular momentum squared after setting the lattice constant to unity. The vectors $\vec a_{\mu}$ are the three translational vectors of the diamond lattice occupied by the Fe atoms. They are given as $\vec a_{1} =\frac{1}{2} (1,1,0)$, $\vec a_{2} =\frac{1}{2} (1,0,1)$, and $\vec a_{3} =\frac{1}{2} (0,1,1)$, which are along the diagonals of the three faces of a cube. To obtain the first ($J_2$) term, we observe that each Fe in a sublattice has twelve second nearest neighbors. For an Fe atom centered at $\vec r_0 = (0,0,0)$, six of these neighbors are positioned at $\vec a_{\mu}, \hspace{0.15cm} \mu =1,2,3$, and their inverses; six others are positioned perpendicular to these directions at vectors $\vec a_{\mu} - \vec a_{\nu}$ with $\mu,\nu = 1,2,3$, and $\mu\neq \nu$. Summing all of these contributions in the continuum limit, one obtains the first term up to an overall total derivative. The last ($J_1$) term can be obtained in a similar fashion by noting that the $J_1$ exchange interaction connects the nearest neighbor, opposite sublattices, as shown in Fig. \ref{DiamondMagnetic} (left). There are four such nearest $J_1$ neighbors for each Fe atom; three lie along the lattice translation vectors ($\vec a_{\mu}$), and one lies within the same primitive cell. The $J_1$ term is then obtained by summing over these contributions in the continuum limit.
 Finally, a biquadratic term (with a coupling constant $K_{\phi}$) for the diamond lattice can be motivated in a manner analogous to the case of a square lattice as was described in the previous paragraph. Fig. \ref{DiamondMagnetic} (left) shows the lattice and magnetic structures of the Fe atoms projected onto the $a$-$b$ plane (i.e. a $c$-axis viewpoint). The red (dark) and green (light) disks denote the two Fe sublattices, and the arrows point in the direction of the spin moments. The topmost (second from top) layer is indicated by a thick (thin) solid disk boundary. These two layers belong to two different sublattices and have antiferromagnetic order in each layer. Even in the presence of a quadratic intersublattice coupling term, the relative orientations of the spins between these two layers are degenerate in the same sense as in Fig \ref{TwoSublattice}. Therefore, the introduction of an intersublattice biquadratic coupling term --- derived by integrating out the short-wavelength, high-energy quantum fluctuations --- will lower this continuous symmetry to an Ising $Z_2$ symmetry. This $Z_2$ symmetry can then be broken at lower temperatures to form a nematic state. For simplicity in the analyses to follow, we ignore longer range exchange couplings, an approximation which is consistent with experiments \cite{Broholm2016}.

We now proceed to decouple the biquadratic term using the Hubbard-Stratonovich transformation. 
At a mean field level, the Hubbard-Stratonovich field ($\equiv \sigma(\vec r)  = \sigma$) plays the role of a nematic order parameter and is proportional to $\langle \phi_1(\vec r) \phi_2(\vec r)\rangle$. A unitary rotation of the fields $\phi_1(\vec r)$ and $\phi_2(\vec r)$ shows that the field $\sigma(\vec r)$ quantifies the degree of a broken $Z_2$ symmetry. The vectors $\vec{\phi}_1$ and $\vec{\phi}_2$ are constrained in this model to lie on a unit sphere, i.e. $\lvert\vec{\phi}_1\rvert^2 =\lvert\vec{\phi}_2\rvert^2=1$. This constraint is imposed through Lagrange multipliers $\lambda_j$ for each of the two fields. Fourier transforming into momentum space and noting that $\phi^*_j(\vec p) = \phi_j(-\vec p)$ (i.e. $\phi_j(\vec r)$ is real), the partition function can be recast into
\begin{figure}[h!]
\includegraphics[width=1.70in,height=1.35in]{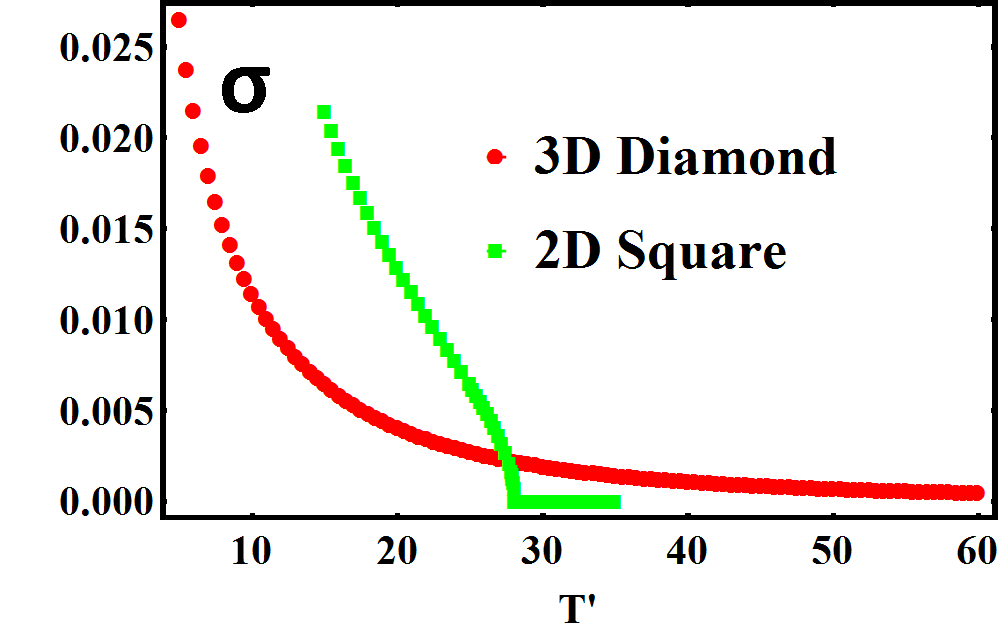}\hfill%
 \includegraphics[width=1.70in,height=1.35in]{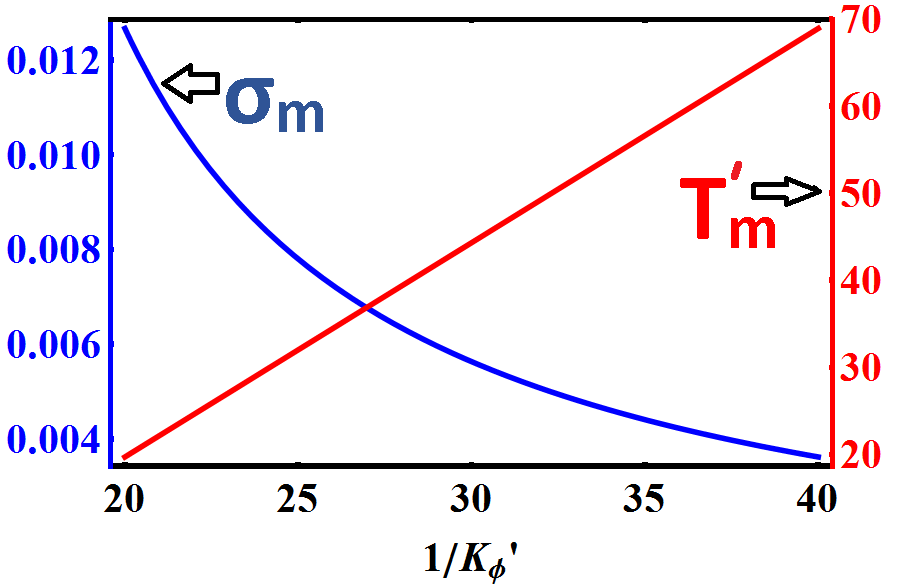}\hfill%
\caption{(Left) Plot of the spin nematic order parameter, $\sigma$, as a function of $T' = T/J_1$ for $\Lambda=2$, $K_{\phi}' = K_{\phi}/J_1= 0.05$. For the sake of comparison, we have also plotted the case of a $2D$ square lattice. (Right) Plots of the magnetic transition temperature, $T_m' = T_m/ J_1$, and the value of the spin nematic order at the magnetic transition temperature, $\sigma_m$, as a function of $1/K_{\phi}'$ for $\Lambda =1$. }\label{SpinNematic}
\end{figure}
\begin{eqnarray}\nonumber
\mathscr{Z} &=& \int \mathscr D\phi_1 \mathscr D \phi_2 \mathscr D \sigma \mathscr D \lambda_1 \mathscr D \lambda_2 \times \exp \left[ \frac{-\beta N}{2} \times \right. \\ 
&& \left. \!\!\!\!\!\!\!\!\!\sum_{\vec p} \left \{ \Phi^{\dagger}(\vec p) M \Phi(\vec p) - 2 T (\lambda_1 + \lambda_2)+ \frac{2 T^2 \sigma^2}{N K_{\phi}}\right\}\right],
\end{eqnarray}
where the matrix elements of the $2\times 2$ matrix $M$ are given by $M_{ii} =  2 \lambda_i T - J_2 \left(\sum_{\vec a_{\mu}} p_{\vec a_{\mu}}^2 + \sum_{\substack{\vec a_{\mu},\vec a_{\nu}\\ \mu<\nu}} p_{\vec a_{\mu} - \vec a_{\nu}}^2 \right)$ for $i=1,2$, and $M_{ij} =  -J_1\left(1+ \sum_{\vec a_{\mu}} p_{\vec a_{\mu}}^2 \right)- 2 T \sigma$ for  $i\neq j$. Here,
 $p_{\vec a_{\mu}} = \vec p\cdot\vec a_{\mu}$, and $\Phi^{\dagger}(p) = \left(\phi_1^*(p), \phi_2^*(p) \right)$. It is easy to check that the $J_2$ terms simply add up to $p^2 = \sum_{i = x,y,z} p_{i}^2$ as was discussed in the preceding paragraph. The $\Phi({\vec p})$ integrals are Gaussian and can be performed easily by standard field theoretic techniques, while the 
remaining functional integrals can be determined by the saddle point approximation. 

The resulting momentum integrals and the simultaneous equations that must be solved for $\lambda_j$ and $\sigma$  are not straightforward; inclusion of the orbital degrees of freedom (see Supplementary Material) only complicates this further, and one must therefore resort to approximations. To do so, we seek hints from experiments \cite{Broholm2016} which provide fits of the data to three different magnetic exchange models. The simplest model assumes that $J_1$ and $J_2$ have opposite signs, and that $\lvert J_1\rvert\ll\lvert J_2\rvert$; this condition implies that we can ignore $J_1$ to the lowest order approximation. By solving the simplified set of equations, however, we find that this approximation does not yield an experimentally consistent variation of the nematic order parameter with temperature. We therefore consider the two other models where $J_1$ is similar in magnitude to $J_2$ and has the same sign. This scenario becomes tedious if the full momentum dependence in $J_1$ is inserted; instead, to allow for analytical transparency, we assume that the $J_1$ term is a constant, independent of momentum. With these approximations, we obtain simultaneous equations for $\lambda$ and $\sigma$ given by (seeking solutions with $\lambda_1=\lambda_2=\lambda$)
\begin{figure}[h!]
\includegraphics[width=1.67in,height=1.4in]{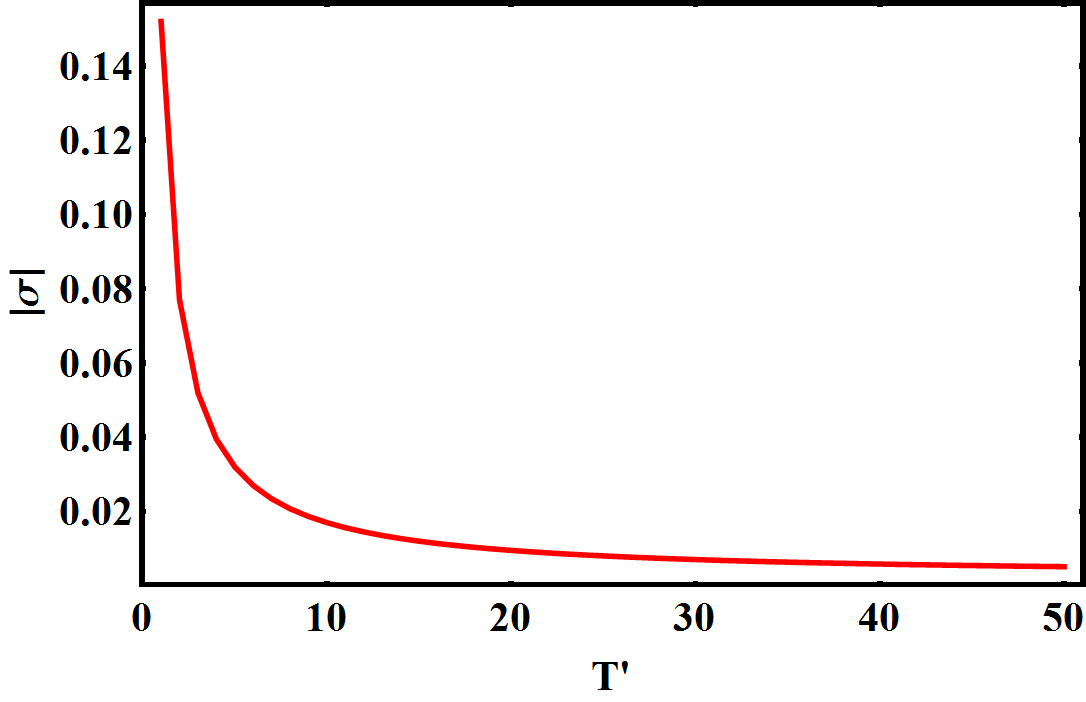}\hfill%
\includegraphics[width=1.67in,height=1.4in]{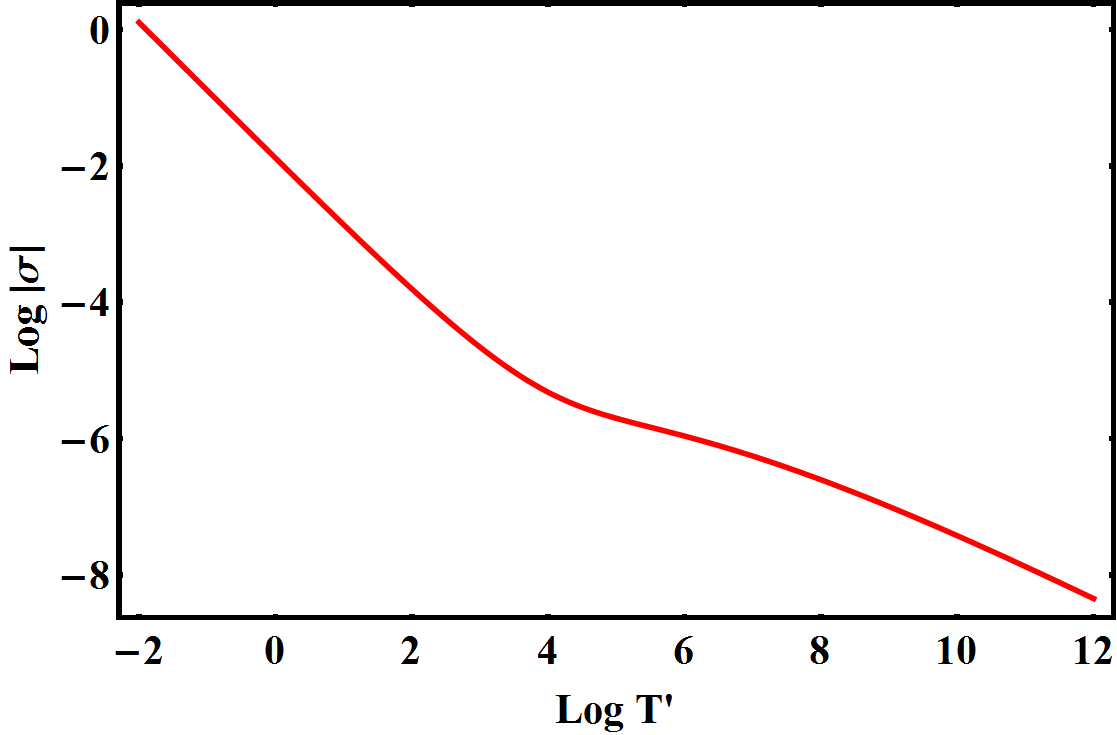}%
\caption{(Left) A plot of the temperature-dependence of the spin nematic order parameter, $\sigma$, obtained by including the full momentum dependence in the $J_1$ term. (Right) Same plot as that on the left but on a log-log scale. The slope in the high $T'$ limit can be shown to be close to $-0.5$ and is confirmed by the numerics above. The parameters chosen are $\Lambda=2$, $K_{\phi}'=0.05$, and $N=3$.}
\label{FullMomentum} 
\end{figure}
\begin{eqnarray}\nonumber
\frac{2 N \pi^2}{T'} &=& -2 \Lambda +  \mathscr{G}_+(\sigma,\lambda,T') +  \mathscr{G}_-(\sigma,\lambda,T'),  \\
-\frac{2 \pi^2 \sigma}{K_{\phi}'} &=& - \mathscr{G}_+(\sigma,\lambda,T') +  \mathscr{G}_-(\sigma,\lambda,T'),
\label{SpinEquations}
\end{eqnarray}
where we have defined $\mathscr{G}_{\pm}(\sigma, \lambda, T') = \sqrt{1\pm 2 T'(\sigma \mp\lambda)} \arctan\left[ \frac{\Lambda}{\sqrt{1\pm 2 T'(\sigma\mp\lambda)}}\right] $. Here, $\Lambda$ is the momentum cutoff and is $\mathscr{O}(1)$ (where the lattice constant is set to unity), $T' = T/J_1$, $K_{\phi}' = K/J_1$, and $J_1 =J_2$. Fig. \ref{SpinNematic}  (left) shows a plot of the spin nematic order parameter, $\sigma$, as a function of $T'$ obtained by numerically solving the above set of equations. Within the aforementioned approximations, $\sigma$ acquires a long tail which slowly vanishes at very large temperatures (compared to the magnetic exchange interactions). It can be checked that at large values of $T'$, the nematic order parameter falls to zero as $T'^{-2}$. The absence of a distinct nematic transition temperature and the presence of a long tail is a result of the three dimensionality of the diamond lattice, indicating that the existence of  multiple sublattices in a cubic system makes it highly susceptible to broken discrete symmetries.
This is unlike the case of a 2D square lattice \cite{Kivelson2008,Kivelson2012} (also shown in Fig. \ref{SpinNematic} (left)) where there is a distinct nematic transition temperature above which the nematic order is zero. These results provide a  possible explanation for the presence of two distinct Fe environments even at high temperatures, as suggested in Ref. \onlinecite{Broholm2016} and also supported by M\"{o}ssbauer \cite{Gibart1976,Kim2008} and far infrared optical spectroscopy \cite{Loidl2015,Haeuseler2002}.

We note that \textit{our qualitative conclusions are robust to the inclusion of the full momentum dependence in the $J_1$ term} as shown in Fig. \ref{FullMomentum} (The accompanying Supplementary Material gives details of the resulting integrals). However, the value of $\gamma$ decreases from 2 to about 0.5 with this inclusion, indicating that the precise value of $\gamma$ could be dependent on the ratio of $J_1$ and $J_2$. That a relatively large NN exchange $J_1$ (comparable to the NNN $J_2$) is needed to obtain experimentally consistent results restricts the possible magnetic models of $\mathrm{FeSc_2S_4}$. (For example, it rules out model 3 in Ref. \onlinecite{Broholm2016}). Finally, our results reveal the presence of a $Z_2$ broken nematic state (which extends up to high temperatures) right above the ordered side of the QCP in the `fan' diagram put forward in Ref. \onlinecite{Schnyder2009}.
 
Next, to obtain the magnetic transition temperature, we need to treat the order parameter field along one of the spin components to be different from those orthogonal to it \cite{Sachdev2007}. In other words, we must integrate out only $N-1$ components and treat the $N$th component as a Lagrange multiplier. Doing so, we obtain the condition for the magnetic transition as $\lambda = \frac{1}{2 T_m'} +  \sigma_m$, where $T_m'$ is the ratio of the magnetic transition temperature to $J_1$, and $\sigma_m$ is the value of the nematic order parameter at the transition temperature. By substituting this condition into Eq. \ref{SpinEquations}, we can solve for $T_m'$ and $\sigma_m$. Fig. \ref{SpinNematic} (right) shows that $T_m'$ grows linearly with inverse $K_{\phi}'$, and for small $K_{\phi}'$, $\sigma_m$ is linearly proportional to $K_{\phi}'$. These conclusions are consistent with our expectations that, depending on their ratio ($K_{\phi}'$), exchange interactions promote magnetic order, while biquadratic couplings favor nematic order. The Supplemental Material describes how this behavior is affected by the presence of  orbital degrees of freedom and the Kugel-Khomskii (KK) type exchange interactions coupling the spins and orbitals. The KK coupling has two qualitatively different consequences: a) \textit{both} the magnetic and orbital ordering temperatures vary with the bi-quadratic interactions and b) the linear dependence of the transition temperatures with $1/K_{\phi}$ $-$ a salient feature of $T_m'$ in the absence of KK interaction (see  Fig. \ref{SpinNematic})$-$ no longer holds good; both $T_m'$ and the orbital equivalent, $T_o'$, now vary sub-linearly. We would also like to point out at this juncture that a solution for the magnetic ordering transition temperature in our model exists only when the signs of $J_1$ and $J_2$ are the same; this reaffirms our previous assertion that we can rule out the magnetic structure of model 3 proposed in Ref. \onlinecite{Broholm2016}. For $\Lambda = 1 $, $J_1 \sim J_2 =0.2\text{ meV}$ (from Ref. \onlinecite{Broholm2016}) and $K_{\phi}' = 0.05$ ($K_{\phi}\ll J_1$), we obtain a magnetic ordering temperature of $T_m =30$ K (compared to the experimental value of 11.8 K). 

To conclude, we modeled the successive breaking of Ising and rotational symmetries in the diamond lattice structure of $\mathrm{FeSc_2S_4}$. We found that, unlike the case of a 2D square lattice, the nematic order for the diamond lattice persists even at high temperatures. Specifically, in the limit $T/J_1\gg1$, the nematic order parameter falls as a power law proportional to $T^{-\gamma}$, $\gamma>0$. This feature indicates that the three-dimensional diamond lattice is unstable toward a $Z_2$ breaking Ising order, and explains the recent observation of two distinct Fe environments in $\mathrm{FeSc_2S_4}$ even at room temperatures. Our theory also restricts the possible magnetic structures and exchange interactions proposed in literature.\\ 
\textit{Acknowledgements:} CS   and   PWP   are   supported by   the   Center   for   Emergent   Superconductivity, a
DOE   Energy   Frontier   Research   Center,   Grant   No. DE-AC0298CH1088.  Partial funding is also provided by the NSF DMR-1461952.   ZL is supported by a scholarship from the Agency of
Science, Technology, and Research.  We thank K. Limtragool for discussions. 

\bibliographystyle{apsrev4-1}
\bibliography{Diamond}
\newpage
\onecolumngrid 

\appendix
\newpage
\section{SUPPLEMENTARY MATERIAL}

\begin{figure}[h!]
\includegraphics[width=2in,height=1.75in]{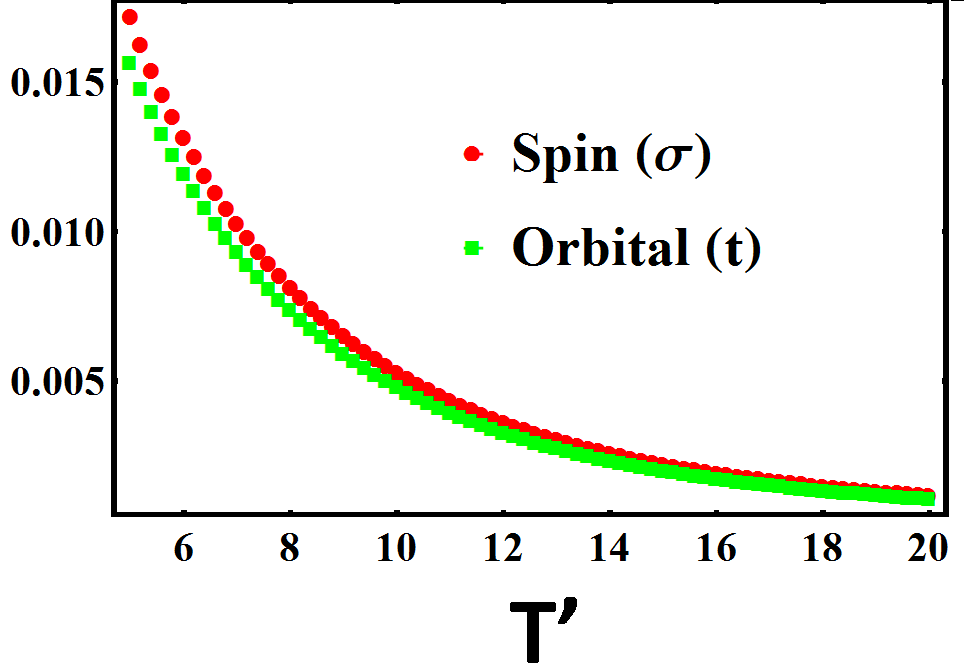}%
 \includegraphics[width=2in,height=1.8in]{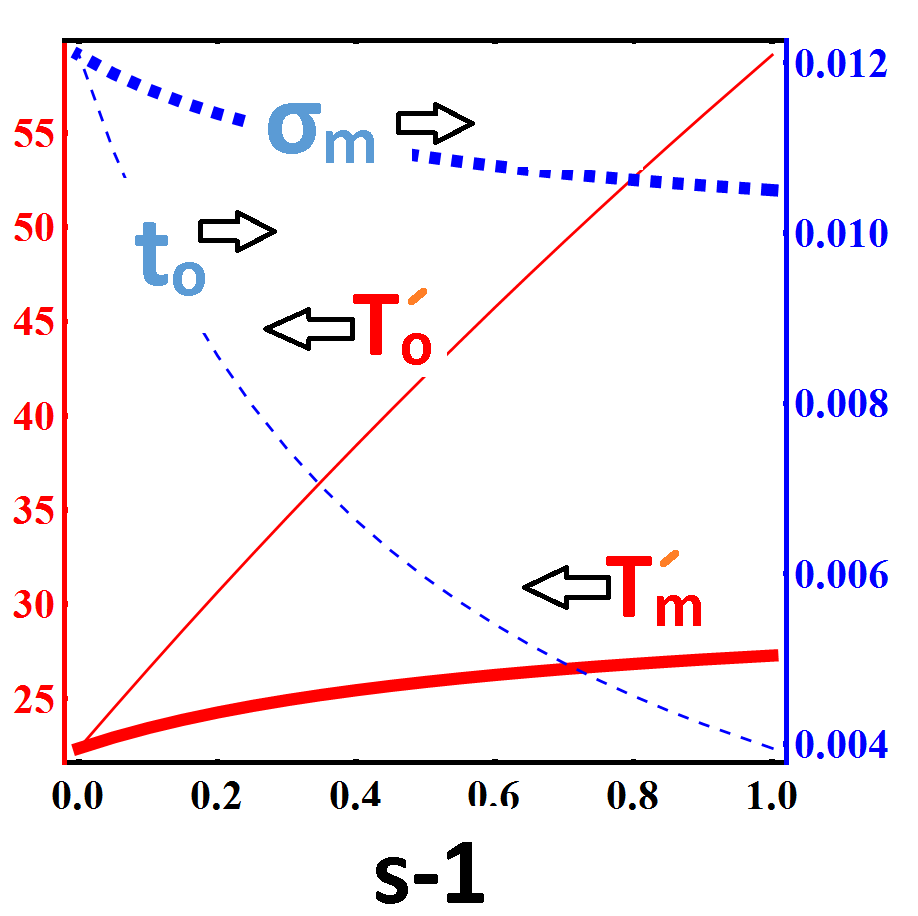}\hfill%
\caption{(Left) Plots for the spin ($\sigma$) and orbital ($t$) nematics as a function of $T' = T/J_1$ for $\Lambda=3$, coupling $\alpha=1$, $K_{\phi}' = 0.02, K_{\tau}' (\equiv K_{\tau}/R_1) = K_{\phi}'/s$ with $s=1.1$. (Right) Combined curves for the magnetic ordering temperature, $T_m'$, orbital ordering temperature, $T_o'$, magnetic nematic at $T_m'$, $\sigma_m$, and orbital nematic at $T_o'$, $t_o$, as a function of $\frac{1}{K_{\phi}'}(s-1)$ (see text) for $\Lambda = 0.9$, $K_{\phi}' = 0.048$, and $\alpha = 0.1$. These various curves are shown by thick solid (red), thin solid (red), thick dashed (blue) and thin dashed (blue) respectively. The arrows mark the axes from which various quantities must be read off.}\label{SpinOrbitalNematic}
\end{figure}

\textbf{\textit{Full momentum dependence in the $J_1$ term:} } In this section, we include the full momentum dependence in the $J_1$ term and show that our conclusions in the main text remain qualitatively the same. The self-consistent equations for the purely magnetic component are given by (all the quantities appearing  have been defined in the main text)

\begin{eqnarray*}
N & = & \int^{\Lambda}\frac{d^{3}p}{\left(2\pi\right)^{3}}\frac{\lambda-\frac{p^{2}}{2T'}}{\left(\lambda-\frac{p^{2}}{2T'}\right)^{2}-\left(\frac{f\left(p\right)}{2T'}+\sigma\right)^{2}},\\
\frac{T'\sigma}{K_{\phi}'} & = & \int^{\Lambda}\frac{d^{3}p}{\left(2\pi\right)^{3}}\frac{\frac{f\left(p\right)}{2T'}+\sigma}{\left(\lambda-\frac{p^{2}}{2T'}\right)^{2}-\left(\frac{f\left(p\right)}{2T'}+\sigma\right)^{2}},
\end{eqnarray*}
where we have defined
\begin{eqnarray*}
f\left(p\right) & = & C+\left(p_{x}+p_{y}\right)^{2}+\left(p_{y}+p_{z}\right)^{2}+\left(p_{z}+p_{x}\right)^{2},
\end{eqnarray*}
and $C$ is a constant. Taking the sum and difference of the above equations and defining $\tilde{\sigma} = 2T'\sigma$, $\tilde{\lambda} =  2T'\lambda$, $\phi_1 = \tilde{\lambda} + \tilde{\sigma}$ and $\phi_2 = \tilde{\lambda} - \tilde{\sigma}$, we obtain
\begin{eqnarray*}
\frac{1}{2T'}\left(N+\frac{\phi_{1}-\phi_{2}}{4K_{\phi}'}\right) & = & \int^{\Lambda}\frac{d^{3}p}{\left(2\pi\right)^{3}}\frac{1}{\phi_{1}-p^{2}-f\left(p\right)},\\
\frac{1}{2T'}\left(N-\frac{\phi_{1}-\phi_{2}}{4K_{\phi}'}\right) & = & \int^{\Lambda}\frac{d^{3}p}{\left(2\pi\right)^{3}}\frac{1}{\phi_{2}-p^{2}+f\left(p\right)}.
\end{eqnarray*}
To diagonalize the denominators of the integrands, we perform the change of variables
$\boldsymbol{p}=U\boldsymbol{k}$ with the orthogonal matrix
\begin{eqnarray*}
U & = & \frac{1}{\sqrt{6}}\left(\begin{array}{ccc}
-2 & 0 & \sqrt{2}\\
1 & -\sqrt{3} & \sqrt{2}\\
1 & \sqrt{3} & \sqrt{2}
\end{array}\right)
\end{eqnarray*}
such that

\begin{eqnarray*}
p^{2}+f\left(p\right) & = & p_{x}^{2}+p_{y}^{2}+p_{z}^{2}+\left(p_{x}+p_{y}\right)^{2}+\left(p_{y}+p_{z}\right)^{2}+\left(p_{z}+p_{x}\right)^{2}+C\\
 & = & \left(\begin{array}{ccc}
p_{x} & p_{y} & p_{z}\end{array}\right)\left(\begin{array}{ccc}
3 & 1 & 1\\
1 & 3 & 1\\
1 & 1 & 3
\end{array}\right)\left(\begin{array}{c}
p_{x}\\
p_{y}\\
p_{z}
\end{array}\right)+C\\
 & = & 2k_{x}^{2}+2k_{y}^{2}+5k_{z}^{2}+C,\\
p^{2}-f\left(p\right) & = & -\left(\begin{array}{ccc}
p_{x} & p_{y} & p_{z}\end{array}\right)\left(\begin{array}{ccc}
3 & 1 & 1\\
1 & 3 & 1\\
1 & 1 & 3
\end{array}\right)\left(\begin{array}{c}
p_{x}\\
p_{y}\\
p_{z}
\end{array}\right)-C\\
 & = & -3k_{z}^{2}-C.
\end{eqnarray*}
Therefore, the self-consistent equations become 
\begin{eqnarray*}
\frac{1}{2T'}\left(N+\frac{\phi_{1}-\phi_{2}}{4K_{\phi}'}\right) & = & \int^{\Lambda}\frac{d^{3}k}{\left(2\pi\right)^{3}}\frac{1}{\phi_{1}-C-2k_{x}^{2}-2k_{y}^{2}-5k_{z}^{2}}\\
 & = & -\frac{1}{8\pi^{2}}\left[\frac{2}{\sqrt{5}}\sqrt{C+2\Lambda^{2}-\phi_{1}}\tan^{-1}\sqrt{\frac{5\Lambda^{2}}{C+2\Lambda^{2}-\phi_{1}}}\right.\\
 &  & \qquad\left.-\frac{2}{\sqrt{5}}\sqrt{C-\phi_{1}}\tan^{-1}\sqrt{\frac{5\Lambda^{2}}{C-\phi_{1}}}+\Lambda\ln\left(1+\frac{2\Lambda^{2}}{C+5\Lambda^{2}-\phi_{1}}\right)\right],\\
\frac{1}{2T'}\left(N-\frac{\phi_{1}-\phi_{2}}{4K_{\phi}'}\right) & = & \int^{\Lambda}\frac{d^{3}k}{\left(2\pi\right)^{3}}\frac{1}{\phi_{2}+C+3k_{z}^{2}}\\
 & = & \frac{\Lambda^{2}}{12\pi^{2}}\sqrt{\frac{3}{\phi_{2}+C}}\tan^{-1}\sqrt{\frac{3\Lambda^{2}}{\phi_{2}+C}}.
\end{eqnarray*}
These equations can then be numerically solved for $\phi_{1},\phi_{2}$,
and consequently $\sigma,\lambda$. The solutions for $\sigma$ are shown in Fig. \ref{FullMomentum} of the main text with $C$ chosen to be equal to unity.\\
\newline\textbf{\textit{Including orbitals:} } In this section, we will incorporate orbital degrees of freedom into the problem by a straightfoward generalization of our analysis of spin moments as done in the main text. We introduce an $O(N)$ orbital vector field, $\vec{\tau}(\vec r)$, on a lattice where the components of each vector at a lattice site denote orbital indices, and the direction of each vector denotes the orbital polarization. At a mathematical level, we essentially have two flavors of $O(N)$ vector fields (like those represented in eq. \ref{SpinLagrangian} of the main text) interacting with each other through a Kugel-Khomskii type spin-orbit coupling term. The new partition function can now be written in a fashion similar to Eq. \ref{SpinPartitionFunction} in the main text and is given as
\begin{eqnarray}
\mathscr{Z}&=&\int \mathscr{D} \vec{\phi}_1 \mathscr{D} \vec{\phi}_2 \mathscr D \vec{\tau}_1 \mathscr D \vec{\tau}_2 \times \exp\left[ - \beta N \int d^3 \vec r \hspace{0.15cm}\left\{ \mathscr{L}_{\phi}\left(\phi_i\right) + \mathscr{L}_{\tau}\left(\tau_i\right) + \mathscr{L}_{\phi\tau}\left(\phi_i, \tau_i\right) \right\}\right]\nonumber,
\end{eqnarray}
where, again, the component index has been suppressed for convenience. Here, $\mathscr{L}_{\phi}$ is the Lagrangian for the spin only sector and $\mathscr{L}_{\tau}$, for the orbital only sector. $\mathscr{L}_{\tau}$ contains   the NN, NNN and biquadratic terms with coupling constants denoted by $R_1$, $R_2$ and $\tilde{K}_{\tau}$ respectively (in this section, for later convenience, we switch the biquadratic couplings to have tildes on top, i.e. $\tilde{K}_{\phi}$, $\tilde{K}_{\tau}$). The last term denotes the interaction between the spins and the orbitals, which we choose to be of a local Kugel-Khomskii type generalization of the biquadratic interaction:
\begin{equation}
\mathscr{L}_{\phi\tau}(\phi_i, \tau_i) = -\tilde{K}_{\phi \tau} \int d^3 \vec r \bigg( \vec{\tau}_1(\vec r).\vec{\tau}_2(\vec r)\bigg) \bigg( \vec{\phi}_1(\vec r).\vec{\phi}_2(\vec r) \bigg).
\end{equation}
The coupling constant for this term is denoted by $\tilde{K}_{\phi\tau}$ and, for simplicity, we assume the number of orbital and spin components to be equal. One can go on to decouple this mixing term using the Hubbard-Stratonovich transform in the spin-spin/orbital-orbital channel; however, we can simplify our analysis by decoupling all the biquadratic terms --- both the pure (spin-spin, orbital-orbital) and mixed terms (spin-orbital) --- with a single transformation. This keeps our formulas tractable while still preserving the non-trivial physics of the coupling term. The transformation takes the form of the integral:
\begin{eqnarray*}
\int_{-\infty}^{\infty} dx dy\hspace{0.15cm}\exp\left[ -(g_1 x^2 + g_2 y^2)  + (x + \alpha y) a   + (\alpha x + y) b \right] &=&\\
&& \!\!\!\!\!\!\!\!\!\!\!\!\!\!\!\!\!\!\!\!\!\!\!\! \frac{\pi}{\sqrt{g_1 g_2}} \exp\left[ \frac{a^2 (g_2 + \alpha^2 g_1) + b^2 (g_1+ \alpha^2 g_2) + 2 \alpha a b (g_1 + g_2) }{4 g_1 g_2} \right],
\end{eqnarray*}
with $g_1,g_2>0$. To put the above transformation into context of our calculation to follow, we set $x=2 N \sigma$, $y = 2 N t$, $g_1 = (4 N^2 \beta K_{\phi})^{-1}$, $g_2 = (4 N^2 \beta K_{\tau})^{-1}$, $a = \phi_1(\vec r) \phi_2(\vec r)$ and $b = \tau_1(\vec r)\tau_2(\vec r)$. We can also read off the relations between ($K_{\phi}, K_{\tau}$) and ($\tilde{K}_{\phi}, \tilde{K}_{\tau}$) from the right hand side of the above transformation by equating
\begin{eqnarray*}
\tilde{K}_{\phi} = \frac{g_2 + \alpha^2 g_1}{4 g_1 g_2};\hspace{1cm}
\tilde{K}_{\tau} = \frac{g_1 + \alpha^2 g_2}{4 g_1 g_2};\hspace{1cm}
\tilde{K}_{\phi\tau}= \frac{2 \alpha (g_1 + g_2)}{4 g_1 g_2},
\end{eqnarray*}
 and then using the relations between $g_i$ and  ($K_{\phi}, K_{\tau}$) given above. We can now use these results and notation to rewrite the partition function as
\begin{eqnarray}
\mathscr{Z} &=& \int \mathscr{D} \vec L \hspace{0.15cm} \exp\left[ \frac{-\beta N}{2} \sum_{\vec p} \left \{ \Phi^{\dagger}(\vec p) M_{\alpha} \Phi(\vec p) + \rho^{\dagger}(\vec p) M_{\alpha}' \rho(\vec p) - 2 T \sum_i (\lambda_i + \lambda_i') +\frac{2 T^2 }{N }\left( \frac{\sigma^2}{K_{\phi}} +\frac{t^2}{K_{\tau}}   \right)\right\}\right],
\end{eqnarray}
where $\mathscr{D}\vec L \equiv \mathscr D\phi_i \mathscr D \tau_i \mathscr D \sigma \mathscr D t \mathscr D \lambda_i \mathscr D \lambda_i'$, $i=1,2$. The constant $K_{\phi}$  is unrelated to that used for our spin only analysis in the main text, but plays a role analogous to it and, hence, we stick to the same notation. The matrix $M_{\alpha}$ includes the coupling terms in the off diagonal elements and is given by
\begin{equation}
M_{\alpha} = 
 \begin{pmatrix}
  2 \lambda_1 T - J_2 \left(\sum_{\vec a_{\mu}} p_{\vec a_{\mu}}^2 + \sum_{\substack{\vec a_{\mu},\vec a_{\nu}\\ \mu<\nu}} p_{\vec a_{\mu} - \vec a_{\nu}}^2 \right) & -J_1\left(1+ \sum_{\vec a_{\mu}} p_{\vec a_{\mu}}^2 \right)- 2 T (\sigma +\alpha t) \\
 -J_1\left(1+ \sum_{\vec a_{\mu}} p_{\vec a_{\mu}}^2 \right)- 2 T (\sigma + \alpha t)  &  2 \lambda_2 T - J_2 \left(\sum_{\vec a_{\mu}} p_{\vec a_{\mu}}^2 + \sum_{\substack{\vec a_{\mu},\vec a_{\nu}\\ \mu<\nu}} p_{\vec a_{\mu} - \vec a_{\nu}}^2\right)
 \end{pmatrix}.
\end{equation} 
The matrix $M_{\alpha}'$ has the same structure as $M_{\alpha}$, but contains the orbital exchange interactions instead. To obtain an explicit form of $M_{\alpha}'$, one only needs to replace the magnetic exchange interactions ($J_i$) with those of their orbital counterparts ($R_i$), $\lambda_i$ with $\lambda_i'$, and swap $\sigma$ with $t$ (i.e. $\sigma \leftrightarrow t$). In the above equations, we have defined  $\rho^{\dagger}(p) \equiv \left(\tau_1^*(p), \tau_2^*(p) \right)$, $\lambda'_i$, the Lagrangian multipliers constraining the orbital vectors fields to have unit magnitude, $t$, the orbital version of the spin nematic ($\sigma$) which we henceforth call the `orbital nematic', and $\alpha$ is a coupling constant proportional to $\tilde{K}_{\phi\tau}$ that mixes the spin and orbital degrees of freedom. We can now integrate out the $\Phi(\vec r)$ and $\rho(\vec r)$ fields and minimize the exponent with respect to $\sigma, t, \lambda, \lambda'$ (like before, we seek solutions with $\lambda_1 = \lambda_2 = \lambda$ and $\lambda_1' = \lambda_2' = \lambda'$) to obtain four self consistent equations given by
\begin{eqnarray}\nonumber
\frac{2 \pi^2 \sigma}{K_{\phi}'}\eta &=& \bigg( \mathscr F\left( g_{\sigma t},\lambda \right) - \mathscr F\left( -g_{\sigma t},\lambda \right)\bigg)+ \alpha  \bigg( \mathscr F\left( g_{t \sigma},\lambda' \right) - \mathscr F\left( -g_{t \sigma},\lambda' \right)\bigg),\\ 
4 \pi^2 N \eta^3 &=& -2 \Lambda \eta +\bigg( \mathscr F\left(- g_{\sigma t},\lambda \right) + \mathscr F\left( g_{\sigma t},\lambda \right)\bigg),
\label{SpinOrbitalEquations}
\end{eqnarray}
 and the two other equations are obtained by interchanging $\sigma \leftrightarrow t$, $\lambda \leftrightarrow \lambda'$ and substituting $K_{\phi}'(\equiv K_{\phi}/J_1) \rightarrow K_{\tau}'(\equiv K_{\tau}/R_1)$ into Eq. \ref{SpinOrbitalEquations} above. In the preceding equations, we have defined $g_{\sigma t} = \left( \eta^2 + \sigma + \alpha t \right)$, $\eta^2 = \frac{1}{2 T'}$, $\mathscr F(x,\lambda) = \sqrt{x-\lambda}\hspace{0.15cm} \arctan\left[\frac{\Lambda \eta}{\sqrt{x-\lambda}}\right]$, and $g_{t\sigma}$ can be obtained by swapping $\sigma\leftrightarrow t$ in $g_{\sigma t}$. Setting $\alpha=0$ gives us two sets of decoupled equations for the $\vec{\Phi}$ and $\vec{\rho}$ fields (similar to Eq. \ref{SpinEquations} in the main text) independent of one another. As will be seen below, the presence of a nonzero $\alpha$ will qualitatively change the behavior of the magnetic ($T_m$) and orbital ordering ($T_o$) temperatures. The conditions determining these temperatures can be derived in a manner similar to that done in the main text. These are given as
\begin{eqnarray}
\lambda &=& \frac{1}{2 T_m'} + \sigma_m + \alpha t_o,\\
\lambda'&=& \frac{1}{2 T_o'} + t_o + \alpha \sigma_m,
\end{eqnarray}
where, $\sigma_m$ and $T_m'$ have been defined in the main text, $t_o$ is the value of the orbital nematic, $t$, at the orbital ordering temperature $T_o$, and the primed quantity $T_o' \equiv T_o/R_1$. Fig \ref{SpinOrbitalNematic} (left) plots both the spin and orbital nematics as a function of $T' = T/J_1$ for fixed values of couplings $K_{\phi}', K_{\tau}'$, $\alpha$, and cutoff $\Lambda$. Their behavior resembles the spin only case (see Fig.\ref{SpinNematic} (left) of the main text) $-$ both $\sigma$ and $t$ acquire a long tails which slowly die for very large temperatures. For large values of $T$ compared to $J_1$ and $R_1$, both the spin and orbital nematic order parameters fall on top of each other go to zero as $T'^{-2}$. At lower values of $T'$, the two order parameters split with $\sigma>t$ if $K_{\phi} > K_{\tau}$ and vice-versa. In both cases, a distinct nematic transition temperature is inherently absent and the role of a non-zero $\alpha$ is to quantitatively enhance both the nematic orders.  \\ \newline
Fig. \ref{SpinOrbitalNematic} (right) shows a plot of the magnetic and orbital ordering temperatures alongside the corresponding values of the nematic order parameters at the transition temperature ($\sigma_m$ and $t_o$). These quantities are plotted as a function of the difference $\left(\frac{1}{K_{\tau}'} - \frac{1}{K_{\phi}'}\right)$, where $K_{\tau}'$ chosen as a scalar multiple of $K_{\phi}'$ (i.e. $K_{\tau}'= K_{\phi}'/s$, $s$ a scalar number). When $s=1$, the coupling constants are equal and both orbital and magnetic ordering occurs at the same temperature with $\sigma_m = t_o$. For $s>1$, these transition temperatures split with $T_o'>T_m'$ while $t_o<\sigma_m$. A nonzero coupling $\alpha$ has two qualitatively different consequences: a) \textit{both} $T_m'$ and $T_o'$ vary with $s$ and b) the linear dependence of the transition temperatures with $1/K_{\phi}$ $-$ a salient feature of $T_m$ when $\alpha=0$ (see  Fig. \ref{SpinNematic} in the main text)$-$ no longer holds good; both $T_m'$ and $T_o'$ now vary sub-linearly with $s$. When $s<1$, on the other hand, the magnitudes of these quantities are reversed. The experimentally relevant scenario seems to be the case with $s>1$ where orbital ordering occurs before magnetic ordering.
\end{document}